\documentclass[12pt,preprint]{aastex}

\usepackage{amsmath}

\begin{document}

\title{Formation of the Double Neutron Star System PSR J1930$-$1852
}
\author{
Yong Shao$^{1,2}$ and Xiang-Dong Li$^{1,2}$}

\affil{$^{1}$Department of Astronomy, Nanjing University, Nanjing 210023, China; shaoyong@nju.edu.cn}

\affil{$^{2}$Key laboratory of Modern Astronomy and Astrophysics (Nanjing University), Ministry of
Education, Nanjing 210023, China; lixd@nju.edu.cn}

\begin{abstract}

The spin period (185 ms) and period derivative ($1.8\times10^{-17}\,\rm s\,s^{-1}$) of the double neutron star (DNS) system PSR J1930$-$1852 recently
discovered indicate that the pulsar was
mildly recycled through the process of Roche-lobe overflow. This system has the
longest orbital period (45 days) of the known DNS systems, and  can be formed from a
helium star-NS binary if the initial mass of the helium star was $ \lesssim 4.0M_{\odot} $; otherwise the helium star would never fill
its Roche-lobe \citep{t15}. At the moment of the supernova explosion, the mass of the helium star
was $ \lesssim3.0M_{\odot} $. We find that the probability distribution of
the velocity kick imparted to the new-born neutron star has a maximum at about
$30 \,\rm km\,s^{-1}$ (and a tail up to $ 260 \,\rm km\,s^{-1}$), indicating that this NS most probably received a low kick velocity at birth.

\end{abstract}


\section{Introduction}

The formation of double neutron star (DNS) systems is believed to be the endpoint
of massive binary evolution \citep[e.g.,][]{bv91,tv06}.
Generally a massive binary first evolves into a high-mass X-ray binary (following the formation
of the first NS) and then evolves through a spiral-in (common-envelope) phase into a
helium star plus NS binary \citep[e.g.,][]{vd73}.
When the helium star explodes as a supernova (SN) to become the second NS, the final binary
may be a DNS system.

Most DNS systems share the characteristics of relatively short spin periods ($ 22.7-104 $ ms), short orbital periods ($0.1-18.8$ days),
and eccentric orbits. However,
PSR J1930$-$1852, a DNS system recently discovered by \citet{s15}, has an
orbital period as long as 45 days and a relatively long spin period of 185 ms.
In this paper, we will discuss the formation history of this system from
helium star-NS binaries.

\section{Formation of PSR J1930$-$1852}

\subsection{Constraints on the parameter space of the helium star-NS binaries}

Evolutionary calculations of helium star-NS binaries have performed by
many authors \citep[e.g.,][]{d02,d03,i03,t15}. During core helium burning
and further burning phases, the helium star loses mass
through stellar winds \citep{h95}.
Of particular interest are low-mass ($ \lesssim3.5M_{\odot} $) helium stars
 since they swell up to large radii during
their late evolution \citep{p71,n84,h86}. The expansion of the helium
star may finally result in the occurrence of Roche-lobe overflow (RLOF) and mass
transfer. The NS can be recycled by accretion of mass and angular
momentum from the helium star companion. If the binary orbit is very wide or
the helium star is too massive, mass transfer via ROLF will not occur prior
to the SN explosion.  For PSR J1930$-$1852, the measured spin period and period derivative
give a relatively weak surface magnetic field of $ 6\times10^{10} $ G, implying
that the pulsar was mildly recycled during the previous mass transfer phase \citep{t15}.

In Fig.~1 we show the plausible distribution of the helium star-NS binary in the final mass ($ M_{\rm He,f}$) of the pre-SN helium star  vs. the pre-SN orbital separation ($a_{0}$) plane.

The pre-SN orbital separation $a_{ 0} $ should lie between the
periastron and apastron of the post-SN orbit with a separation of $a$ \citep{f75}, i.e.,
\begin{equation}
a(1-e)\leq a_{\rm 0} \leq a(1+e),
\end{equation}
where $ e $ is the post-SN eccentricity. In Fig.~1 we indicate the maximum
and minimum pre-SN orbital separations with the two black solid lines by adopting
$a=73R_{\odot} $ and $e=0.4$.

The fact that PSR J1930$-$1852 has experienced mass transfer in  a wide orbit indicates that $ M_{\rm He,f} \lesssim 3M_{\odot}$ as indicated by the blue dashed line in Fig.~1, 
which distinguishes whether the helium star can evolve to fill its RL.
For a system with a wide orbit at the time of the SN explosion, the preceding mass transfer
should have decreased the orbital separation, so the initial separation can have
been wider as indicated by the blue dotted line in Fig.~1. The calculated results of \citet[][see their Table~1]{t15} 
were used to plot
these two lines. From this figure
we can estimate that the initial mass of the helium star is $ \lesssim4M_{\odot} $.

\subsection{Asymmetric SN explosions and NS kicks}

When the helium star exploded as a SN, a kick velocity $ V_{\rm k} $ was imparted to the second  NS.
The orientation of the kick velocity is controlled by the angles $ \theta $ and $ \phi $, as shown in Fig.~2. Here $\theta$ is the angle between the kick velocity and the pre-SN orbital velocity $V_0$, and $ \phi $ is the positional angle of the kick velocity with respect to the orbital plane.
Thus the relation between $a_{0} $ and $ a $ is given by \citep{h83,d03},
\begin{equation}
\frac{a_{0} }{a} = 2-\frac{M_{0}}{M}(1+\nu^{2}+2\nu\cos\theta),
\end{equation}
where $M_{0} =   M_{1}+ M_{\rm He,f} $, $ M = M_{1}+ M_{2}$, $ \nu= V_{\rm k}/V_{0}$, and $ V_{0} = (GM_{0}/a_{0})^{1/2} $; $ M_{1} $ and $ M_{2} $ are the pulsar mass and the second NS mass respectively. We assume $ M_{1}= M_{2} = 1.3M_{\odot}$
in the following calculation. The eccentricity of the post-SN orbit can be written as \citep{h83,d03}
\begin{equation}
1-e^{2} = \frac{a_{\rm 0} }{a}\frac{M_{0}}{M}[1+2\nu\cos\theta
+\nu^{2}(\cos^{2}\theta+\sin^{2}\theta\sin^{2}\phi)].
\end{equation}
In Eqs.\,(2) and (3) there are five variables, that is, $ V_{\rm k} $,
$ M_{\rm He,f}$, $ a_{0} $, $ \theta $, and $ \phi $.
We can derive useful information on the distribution of $ V_{\rm k} $ if the values of
$ M_{\rm He,f}$ and $ a_{0} $ are reasonably constrained.

From Eqs.~(2) and (3), the condition of $ \sin^{2}\phi =1 $ always gives the solution
$ a_{0} = a(1\pm e) $, independent of the magnitude of the
$ V_{\rm k} $. This result corresponds to the lower and upper limits of the orbital separation,
as given by Eq.~(1). For the case of $ \sin^{2}\phi =0 $, we can derive two boundary lines
for any specific value of  $ V_{\rm k} $. In Fig.~1, the orange and green lines correspond to the
$ V_{\rm k} = 30$ and $ 50 \,\rm km\,s^{-1} $, respectively. In a special situation that a new born
NS has no kick ($ V_{\rm k} = 0 $), there are two solutions with $ M_{\rm He,f} =2.34M_{\odot} $,
$ a_{0} =43.8R_{\odot} $ and $ M_{\rm He,f} =0.26M_{\odot} $, $ a_{0} =102.2R_{\odot} $.

For any fixed value of $ V_{\rm k} $, the possible values of $ M_{\rm He,f}$ and $ a_{0} $
are confined by the conditions of $ \sin^{2}\phi =0 $ and $ \sin^{2}\phi =1 $. As an
illustration, Fig.~3 shows the allowed distribution of  $ M_{\rm He,f}$ and $ a_{0} $
with $ V_{\rm k} =40 \,\rm km\,s^{-1}$. The thick green and red lines show the solutions
in the limits of  $ \sin^{2}\phi =0 $ and 1, respectively. We set 1000 random values for both
$ \theta $ and $ \phi $ in the interval of $ 0-\pi $, and plot the solutions
in Fig.~3 as triangles.

\citet{h86} suggested that the threshold mass for a single helium star to produce a NS is $ \sim2.2M_{\odot} $. If the helium star is in a binary, the threshold mass
is determined by the CO or ONeMg core mass \citep{n84}.
Here we assume that the minimal mass of a pre-SN helium star is $ 1.4M_{\odot} $, which
is compatible with the calculated results of \citet{t15}, while
the pre-SN maximal mass is $ \sim3M_{\odot} $ (see Section 2.1).
We ran millions of numerical
calculations with randomly distributed angles $ \theta $ and $ \phi $. Using the limits of $ 1.4M_{\odot}
\leq M_{\rm He,f} \leq 3M_{\odot}$ and Eq.~(1), we derived the distribution of the kick
velocities. Figure~4 shows the normalized and accumulated distributions of $ V_{\rm k} $
with the black and red curves, respectively. We find that the $ V_{\rm k}$-distribution has a peak
at $ \sim30 \,\rm km\,s^{-1}$ with a maximum value $ \sim260 \,\rm km\,s^{-1} $.
For $ V_{\rm k} \leq 50$ and $ 100 \,\rm km\,s^{-1} $, the generated probabilities in our calculations
are 0.63 and
0.77, respectively. In order to reproduce the observed parameters of PSR J1930$-$1852, the higher
the value of $ V_{\rm k} $, the more restricted the orientation of the NS kick.
We therefore see that the most likely value of $ V_{\rm k} $ was low, of the order
of $30 \,\rm km\,s^{-1}$.

\section{Summary}

Our analysis reveals  the following results.

(i) To satisfy the condition that there was ROLF-mass transfer in the progenitor
helium star-NS binary with
a wide orbit, the mass  of the pre-SN helium star  $ M_{\rm He,f}$ should be $\lesssim3.0M_{\odot} $
and correspondingly the  initial mass $ \lesssim 4M_{\odot}$.

(ii) For randomly orientated NS kicks, the most likely values of  the kick
velocities are low with a peak at $ \sim30 \,\rm km\,s^{-1}$.

Recently \citet{b15} demonstrated that the second collapse in the majority of DNS systems
involved small mass ejection ($ \Delta M\lesssim 0.5M_{\odot} $) and a low kick
velocity ($V_{\rm k} \lesssim 30\,\rm km\,s^{-1}$), which may be related to the electron-capture SNe
as suggested by \citet{p04} and \citet{v04}.
For the helium star-NS progenitor of PSR J1930$-$1852, both the helium star's mass and the
kick velocity are relative low, which are indeed likely to correspond with
the second NS having originated from an electron-capture SN.
Still the formation channel of core-collapse SN can not be excluded, and further observations
can help settle this problem.

\acknowledgements
This work was supported by the Natural Science Foundation of China
under grant numbers 11133001, 11203009 and 11333004, the Strategic
Priority Research Program of CAS (under grant number XDB09000000).


\clearpage

\begin{figure}

\centerline{\includegraphics[scale=0.8]{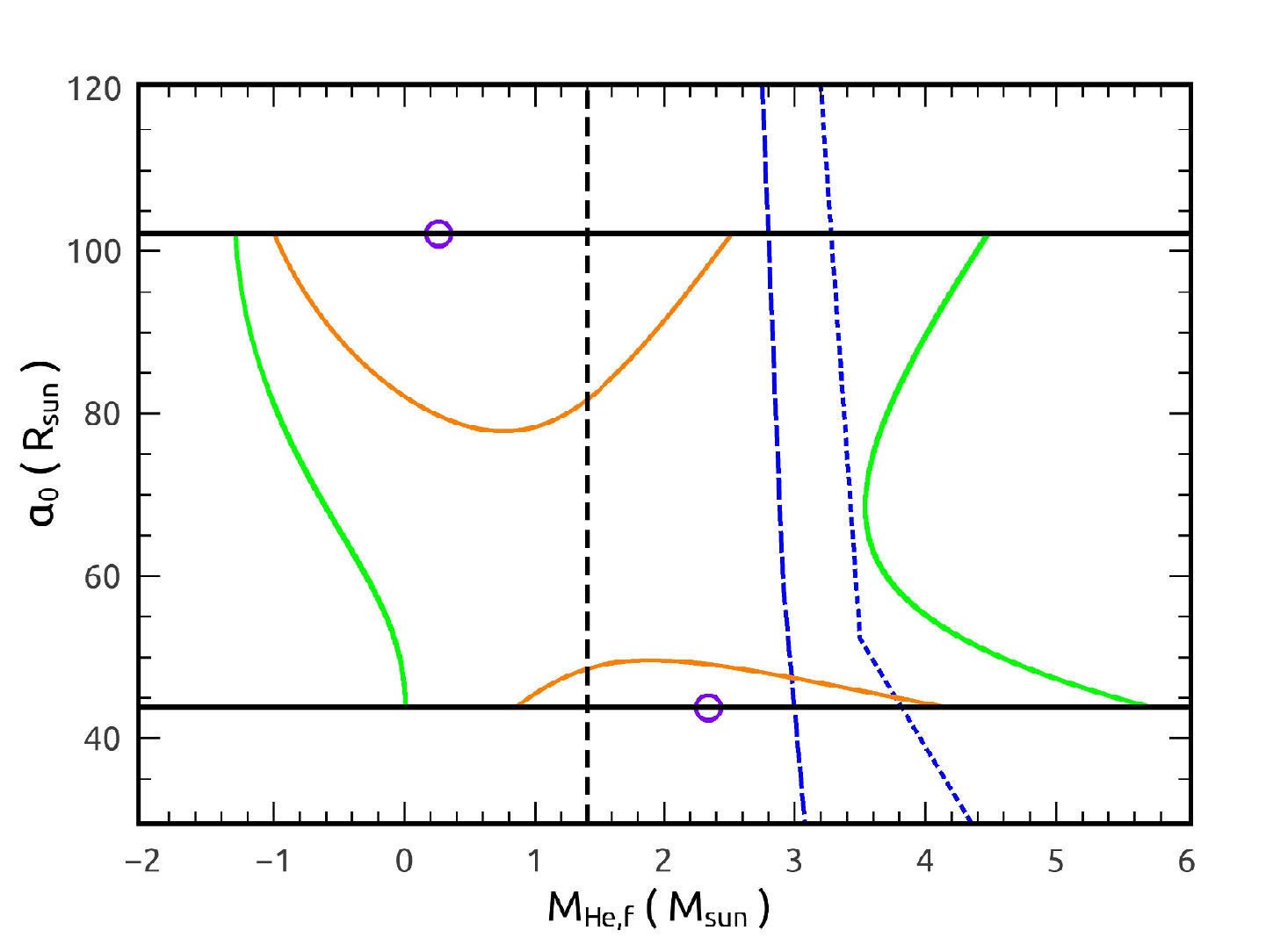}}
\caption{The allowed distribution of the pre-SN helium star mass $ M_{\rm He,f}$
and the orbital separation $a_{0} $. The blue dashed line distinguishes whether the helium star can
fill its RL prior to the SN explosion, and the blue dotted line denotes the corresponding relation
at the beginning of binary evolution with an unevolved helium star \citep{t15}.
The pre-SN orbital separation of PSR J1930$-$1852 is limited by
the two black solid lines. The orange and green lines correspond to the solutions of Eqs. (2) and (3)
when $ \sin^{2}\phi=0 $ in the cases of $ V_{\rm k} = 30$
and $ 50 \,\rm km\,s^{-1} $, respectively. The two circles mark the positions when no kick is impacted
to the second NS. The minimal mass of $ 1.4M_{\odot} $ for the helium star
is presented with the dashed black line.
  \label{figure1}}

\end{figure}

\begin{figure}

\centerline{\includegraphics[scale=0.8]{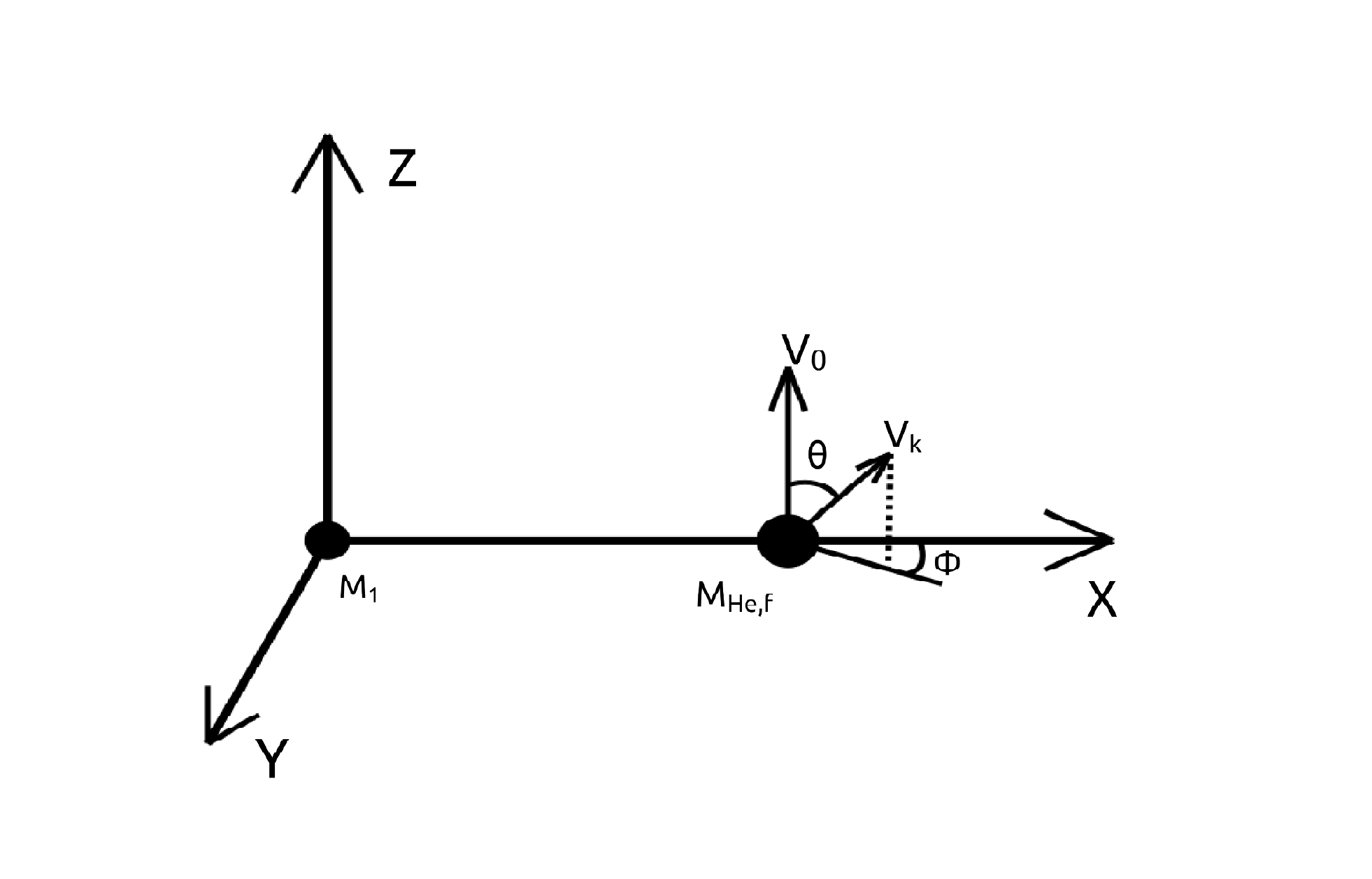}}
\caption{The orientation of the kick velocity  relative to the original orbital
velocity and the orbital plane of a helium star-NS binary.
 \label{figure2}}

\end{figure}

\begin{figure}

\centerline{\includegraphics[scale=0.8]{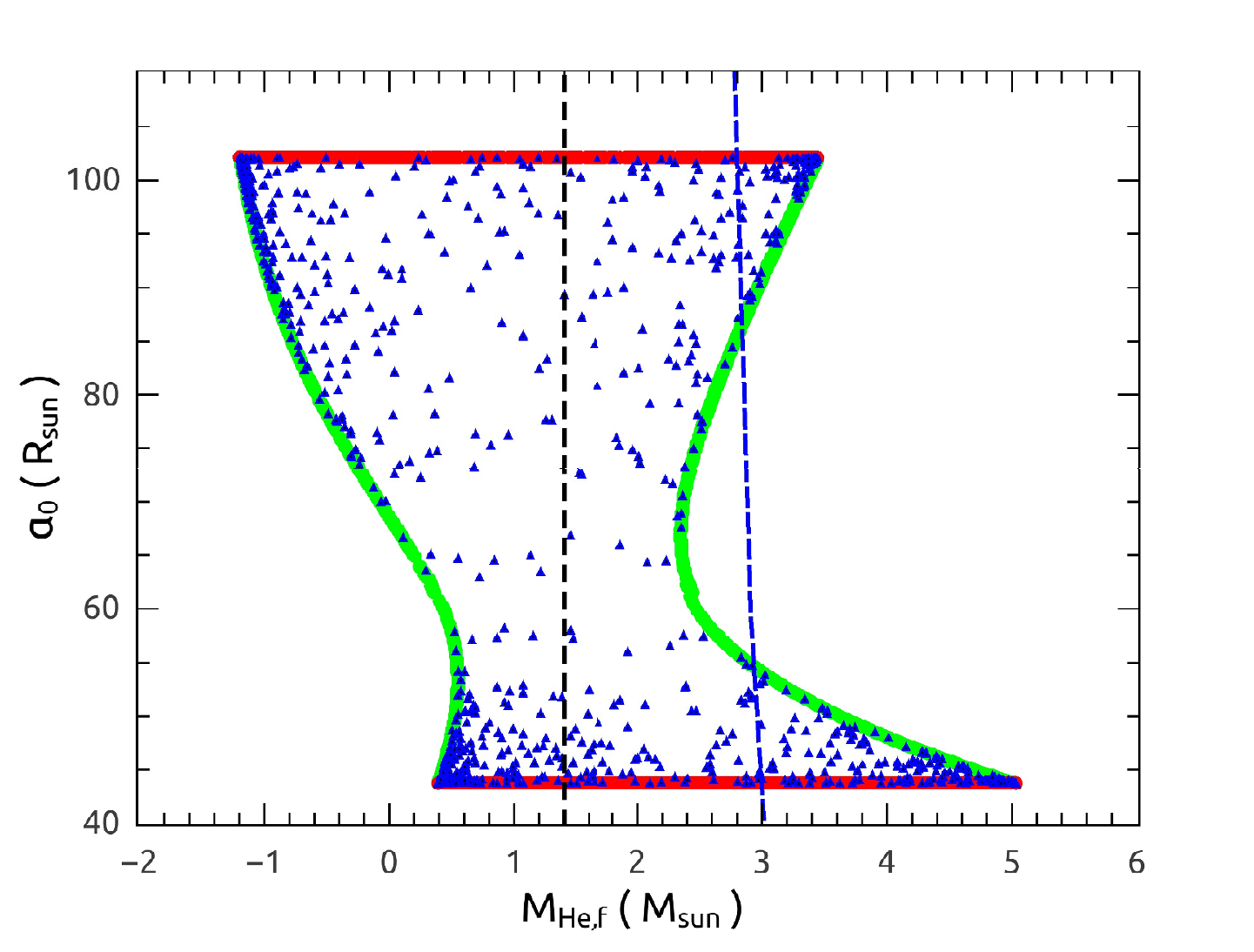}}
\caption{The distribution (blue triangles) of the calculated $ M_{\rm He,f}$ and $a_{0} $
from Eqs.~(2) and (3) with randomly distributed angles $ \theta $ and $ \phi $. Here $ V_{\rm k} $ is set to be
$40 \,\rm km\,s^{-1} $. The thick green and red lines correspond to the cases of $ \sin^{2}\phi =0 $
and 1, respectively. The thin dashed lines have the same meanings as in Fig.~1.
\label{figure3}}

\end{figure}

\begin{figure}

\centerline{\includegraphics[scale=0.8]{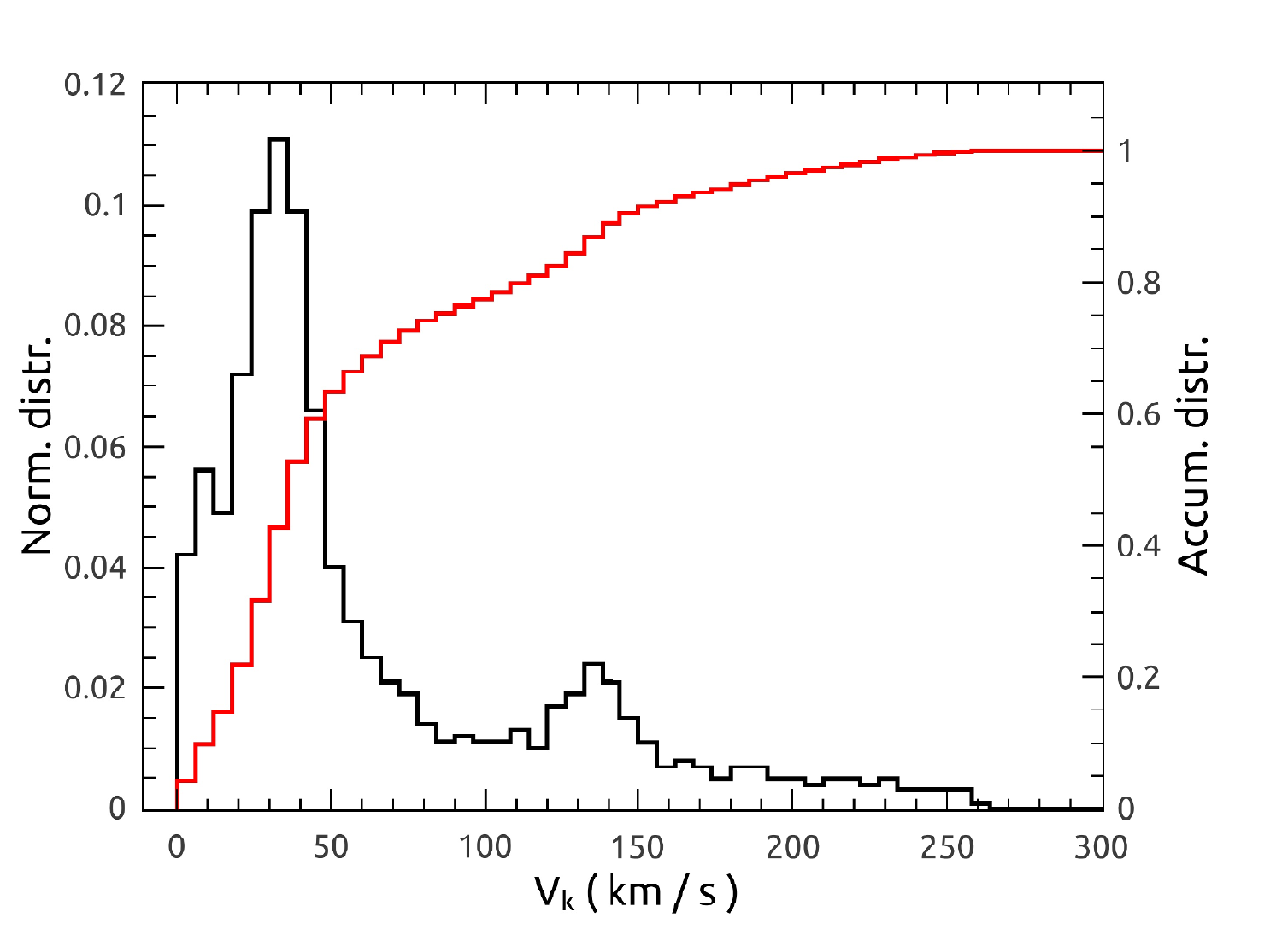}}
\caption{The normalized (black line) and accumulated (red curve) distributions for the calculated kick velocities.
\label{figure4}}

\end{figure}

\end{document}